\documentclass[showpacs,aps]{revtex4}
\usepackage{graphicx,color,textcomp}
\usepackage{amsmath}

\begin{document}

\title{Convection induced nonlinear-symmetry-breaking
in Wave Mixing}

\author{Roberta Zambrini\email{roberta@phys.strath.ac.uk}}

\email{roberta@phys.strath.ac.uk}
\affiliation{Department of Physics, University of Strathclyde Glasgow,
G4 0NG, UK.}

\author{Maxi San Miguel}
\affiliation{Instituto Mediterr{\'a}neo de Estudios Avanzados,
IMEDEA (CSIC-UIB), Campus Universitat Illes Balears, E-07071 Palma
de Mallorca, Spain.}

\author{C\'eline Durniak}
\author{Majid Taki}
\affiliation{Laboratoire de Physique des Lasers, Atomes et
Mol\'ecules, CNRS UMR 8523, Centre d'Etudes et de Recherches Lasers et
Applications, Universit\'e des Sciences et Technologies de Lille,
F-59655 Villeneuve d'Ascq Cedex, France.}

\date{\today}
\begin{abstract}

We show that the combined action of diffraction and convection
(walk-off)  in wave mixing processes leads to a
nonlinear-symmetry-breaking in the generated traveling waves. The
dynamics  near to threshold is reduced to a Ginzburg-Landau model,
showing  an original dependence of the nonlinear self-coupling term
on the convection. Analytical expressions of the intensity and the
velocity of traveling waves emphasize the utmost importance of
convection in this phenomenon. These predictions are in excellent
agreement with the numerical solutions of the full dynamical model.

\end{abstract}

 \maketitle

There is currently a considerable interest in understanding the role
of $convection$ in pattern forming systems
\cite{Cross93,Lugiato0,Rosanov} in such diverse fields as
hydrodynamics \cite{Hydro}, plasma physics \cite{Plasma}, traffic
flow \cite{Mitarai2000}, crystal growth \cite{Israeli}, and
nonlinear optics \cite{Maxi97}. The most important and common
result, in these studies, is that the pattern selection in spatially
extended systems is dramatically affected by the breaking of the
reflection symmetry ($\vec{r}\rightarrow -\vec{r}$) due to the
presence of convection. 
This is a $linear$ symmetry breaking where convection terms are
generally considered to have the only effect to induce a
traveling character to selected patterns, and to induce a
peculiar regime of convective instability. It has been largely
studied how the existence, the type and the dynamics of the
pattern selected are closely related to the linear transition
from convective instability (propagation overcomes amplification
of perturbations) to absolute instability (amplification
dominates).

In contrast, in this paper, we discuss an unforeseen 
effect of convection in the dynamics of spatially
extended systems that does not rely on such a transition.  Here, we
show how convection, that is a linear phenomenon, actually modifies
the intrinsic nonlinearity of the system. More precisely, we show
that convection affects the nonlinear modes interaction at onset of
the instability leading to a {\em nonlinear symmetry breaking} in the
generated, otherwise symmetrical, traveling waves. Our analytical
description of this mechanism, based on the amplitude equation of
the degenerate optical parametric oscillator, demonstrates an
original dependence of the nonlinear self-coupling term upon
convection.
We consider a degenerate optical parametric oscillator (DOPO) because
convection (walk-off) arises naturally from the birefringence of the
crystal that composes this device. Moreover these devices are at the
basis of interesting quantum phenomena, stemming from their quadratic
non-linearity, as for instance entanglement between off-axis modes
\cite{robyEPJD}.  It has been shown that the walk-off strongly
influences such twin beams correlations: in the convective  regime
(induced by walk-off) the entanglement is destroyed by macroscopic
amplification of quantum noise \cite{PRAconv}. Increasing the pump
intensity,  an absolutely stable traveling pattern arises in the
signal and the entanglement is restored. Still, important  walk-off
effects are observed, as  one of the twin beams is more intense and it
fluctuates more than the opposite one \cite{PRAtwin}. Although, we
present our investigations in the context of optics, we believe that
our result is generic for spatially extended systems with convection
and characterizes the key role of convection in the nonlinear dynamics
of such systems.

We start from the description of a type I phase-matched DOPO
in the mean-field approximation
including diffraction and walk-off \cite{Marco+Ward98}
\begin{eqnarray}
\partial_{t}A_{p} &=& \gamma _{p}[-(1+i\Delta_{p})A_{p}+ia_{p}{\nabla}
^2_{\perp }A_{p}-A_{s}^{2}+E_0]  \label{eqdopo} \\
\partial _{t}A_{s}&=& \gamma _{s}[-(1+i\Delta _{s})A_{s}+ia_{s} {\nabla}
^2_{\perp }A_{s}+A_{p}A_{s}^{*}-\alpha _{s}\partial _{x}A_{s}] \nonumber
\end{eqnarray}
\noindent where $A_{p}$ and $A_{s}$ are the normalized slowly
varying envelopes for pump and signal fields, respectively. The
parameters $\Delta _{p,s}$, $\gamma _{p,s}$, and $a_{p,s}$ are the
detunings, the cavity decay rates and the diffraction coefficients,
respectively. $E_0$ is the normalized external pump and $\alpha
_{s}$ is the signal walk-off coefficient that characterizes
convection in this system. We stress that the walk-off cannot be
eliminated by a change of reference frame, being relative between
pump and signal.

Both convective and absolute instabilities  have been reported for
the stationary  solution $A_p=E_0/(1+i\Delta_p)$, $A_s=0$
\cite{Marco+Ward98}. We just recall that for the case $\Delta_s < 0
$ and $\Delta_s \Delta_p -1 < 0$, which we are interested in,
degenerate OPOs exhibit a supercritical bifurcation at
$E_0^c=E_0/\sqrt{1+\Delta _{p}^{2}}$. This is the linear threshold
at which stationary homogeneous solutions become unstable to
traveling wave perturbations with wave vectors $\vec{k}=(k_x,k_y)$
with modulus $k=k_c=\sqrt{-\Delta_s/a_s}$ and frequency $\omega_c=
-\gamma_s \alpha_s k_c$. Under periodic boundary conditions the
convective instability is suppressed and the traveling rolls arising
at the signal generation threshold are absolutely stable.

To study the role of convection in the nonlinear symmetry breaking
of the generated traveling waves, and to keep mathematics as simple
as possible, we perform the reduction of the model (\ref{eqdopo})
into a single Ginzburg-Landau (GL) equation valid close to threshold
of the DOPO emission. In the sequel we set $\mu _{p}=E_0/(1+i\Delta
_{p})$ with $\mu =|\mu _{p}|=E_0/ \sqrt{1+\Delta _{p}^{2}}$, and
$B=A_{p}-\mu _{c}$ where $\mu _{c}=1$. We expand the signal as
$A_{s}=\varepsilon A^{(1)}+\varepsilon ^{2}A^{(2)}+\varepsilon
^{3}A^{(3)}+...$ with a similar expansion of the pump. The small
parameter $\varepsilon $ measures the distance from the DOPO
emission threshold: $\varepsilon ^{2}=\mu -\mu_{c}$.
Setting $A^{(1)}=A\exp i(\omega
_{c}T_{0}+k_{c}X_{0})+A^{\ast }\exp -i(\omega _{c}T_{0}+k_{c}X_{0})$
and applying the solvability condition at order $O(\varepsilon
^{3})$ we get the following amplitude equation, which describes the
evolution of the signal written in the scaled time $\tau =\gamma
_{s}t$, and in the variable $ S=\varepsilon A $
\begin{equation}
\partial _{\tau }S+\alpha _{s}\partial _{x}S=(\mu -1)S-2a_{s}\Delta
_{s}\mathcal{L}_{xy}^{2}S-\beta \text{ }\left| S\right| ^{2}S  \label{eqGL}
\end{equation}
\noindent with
\begin{equation}
\beta =\frac{2}{1+\Delta _{p}^{2}}+\frac{C^{2}+1+D^{2}+iD(
C^{2}-1-D^{2}) }{ [1+(C+D)^{2}][1+(C-D)^{2}] }  \label{eqbeta}
\end{equation}
\noindent We have set $\mathcal{L}_{xy}=(\partial _{x}+\partial
_{y}^{2}/{2ik_{c}})$ in Eq.~(\ref{eqGL}) and $C=\Delta
_{p}+4a_{p}k_{c}^{2}$ and $D=2\omega _{c}/\gamma_{p}=-2\alpha
_{s}(\gamma _{s}/\gamma _{p})k_{c}$ in the expression of the
nonlinear self-coupling coefficient $\beta$. In absence of
convection ($\alpha_s=0$), the parameter $D$ vanishes and the
expression of $\beta $ greatly simplifies to  $\beta ={2}/{(1+\Delta
_{p}^{2})}+{1}/{(1+C^{2})}$ and stationary rolls arise in the signal
profile \cite{Oppo}. We show here that the presence
of convection drastically affects the pattern formation mechanism
with respect to both linear and, more importantly, nonlinear
dynamics. Indeed, the most important result is that the parameter
$D$, that characterizes convection, strongly modifies the nonlinear
self-coupling term $\beta$. It affects the saturation term
[$\mathrm{Re}(\beta)$] and induces intrinsic nonlinear phase
modulations [$\mathrm{Im}(\beta)$]. This result is in contrast with
almost all previous studies of model equations describing the
near-threshold dynamics such as the Ginzburg-Landau or
Swift-Hohenberg equations where the convection only yields to the
propagation term of Eq.~(\ref{eqGL}). No study has yet been
reported, to our best knowledge, on any dependence of the nonlinear
coefficient $\beta$ upon convection ($\alpha_s$).

At this stage one has to notice that the presence of
convection via the non zero $\mathrm{Im}(\beta )$ breaks the
well-known variational form of the GL Eq.~(\ref{eqGL})
 in its one-dimensional version (where 
 $\mathcal{L}_{xy}=\partial _{x}$), since all the
remaining coefficients are real \cite{Fauve90}. As a consequence 
Eq.~(\ref{eqGL}), obviously, cannot exhibit stationary homogeneous
solutions (or stationary rolls). Moreover, the non variational
effect leads to (i) a symmetry breaking in the opposite traveling
waves and (ii) an excess velocity (with respect to convection
velocity) in these waves stemming from the nonlinear frequency
modulation. Both points are analytically characterized below.

Let us find the solutions of Eq.~(\ref{eqGL}), corresponding to the
nonlinear saturated selected modes,
 in the form $S_{st}=S_{0}\exp i(\Omega \tau +kx)$.
 They read $\left| S_{0}\right| ^{2}=(\mu -1+2a_{s}\Delta _{s}k^{2})/
\mathop{\rm Re}(\beta )$ and $\Omega =-\alpha
_{s}k-\mathrm{Im}(\beta )|S_{0}|^{2}$, and represent the leading
contribution to the fundamental modes ($\pm k_c$).
This leading contribution is not sufficient 
 since the total intensity of each mode is now fixed
during their nonlinear interaction induced by the convection. For
our purpose to characterize the nonlinear symmetry breaking, we
need to take into account the contributions up to the third order in
$\varepsilon.$ This can be achieved by solving the hierarchy of the
inhomogeneous linear problems, at each order in $\varepsilon$, by
means of Fredholm alternative. After lengthy but straightforward
calculations, we get the solution
\begin{eqnarray}
A_{s} &=&[1+(1/2)F_{3}\left| S_{st}\right| ^{2}] S_{st}\exp i(\omega
_{c}t+k_{c}x)  \nonumber \\
&+&[1-(1/2)F_{3}^{*}\left| S_{st}\right| ^{2}]S_{st}^{*}\exp -i(\omega
_{c}t+k_{c}x)  \label{eqAs}
\end{eqnarray}
where $F_{3}$ is defined as $\mathop{\rm Re}(F_{3})=2CD/Den$ and $
\mathop{\rm Im}(F_{3})=2\Delta _{p}/(1+\Delta
_{p}^{2})+C(1+C^{2}-D^{2})/Den$ with
$Den=[1+(C+D)^{2}][1+(C-D)^{2}]$. Note that the spatial
modulations of these traveling waves are not relevant here and have
been neglected (i.e. $k=0$) in writing the above solution that is
still composed of two asymmetric nonlinear traveling waves. The
nonlinear symmetry breaking depends on the set of parameters in
which the DOPO operates via the ratio between the intensities (i.e.
$R^2=| A_{s}^2(k_{c})| /| A^2_{s}(-k_{c})|=I(k^c)/I(-k^c) $) of
the two transverse modes of the signal [Eq.~(\ref{eqAs})]. This
ratio has the explicit form
\begin{equation}
R^{2}=1+\frac{\mathop{\rm Re}(F_{3})}{\mathop{\rm Re}(\beta )}\frac{2}{
u_{1}^{2}+u_{2}^{2}}(\mu -1)  \label{eqR2}
\end{equation}
\noindent with $u_{1}=1-[\mathop{\rm Re}(F_{3})/2\mathop{\rm Re}(\beta)](\mu -1)$ and $u_{2}=[\mathop{\rm Im}(F_{3})/2\mathop{\rm Re}(\beta )](\mu-1)$.

\begin{figure}[tbp]
\includegraphics[width=8cm]{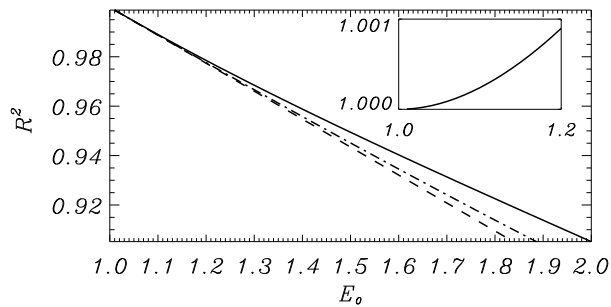}
\caption{\protect Dependence of $R^2$ on the pump
parameter $E_0$ above threshold ($E_0^c=1$). Numerical data
obtained from integration of Model (\ref{eqdopo}) (continuous line)
compared with predictions of Eq.~(\ref{eqR2}) (dashed-dotted line) and
Eq.~(\ref{eqR2CD}) (dashed line). The insert shows the ratio between
the continuous and the dashed-dotted lines.
The parameters are: $\gamma_p=\gamma_s=1$, $\Delta_p=0$,
$\Delta_s=-1$, $a_p=0.25$, $a_s=0.5$, and $\alpha_s=0.25$.}
\label{fig02}
\end{figure}

This is the main analytical result. It allows a quantitative
characterization of the nonlinear convection effects. Equation
(\ref{eqR2}) emphasizes the coupling between convection and the
distance from threshold. In the absence of convection $R^2=1$, the two
transverse modes have the same intensity and the amplitude equation
exhibits stationary rolls. The presence of convection greatly
complicates the expression of $R^2$. However, near threshold ($\mu
\gtrsim 1$), the ratio of intensities $R^{2}$, up to the leading order in $\mu -1$, is
\begin{equation}
R^{2}\simeq 1+4\frac{C(1+\Delta _{p}^{2})}{2Den+(1+\Delta
_{p}^{2})(1+C^{2}+D^{2})}D(\mu -1)\text{ }  \label{eqR2CD}
\end{equation}
\noindent Note that $R^2-1$ is an odd function of $\alpha_s$, reflecting the
importance of the sign of the velocity convection. Therefore, the choice of the
convection direction ($\pm \alpha_s$) can be useful to select one of the two
modes by enhancing its parametric gain. Figure \ref{fig02} shows a typical
variation of $R^2$ upon the physical pump amplitude $E_0=\mu
\sqrt{1+\Delta_p^2}$.
 We find a very good agreement between the analytical ratio
$R^2$  and the numerical simulations of the
Eq.~(\ref{eqdopo}). In order to set the validity range of our
predictions we have plotted the results obtained by increasing the
pump till twice the threshold. Even for pump values $20\% 
$ above threshold the agreement is within  $1$\textperthousand (see insert
in Fig.~\ref{fig02}).

\begin{figure}[tbp]
\begin{center}
\includegraphics[width=8cm]{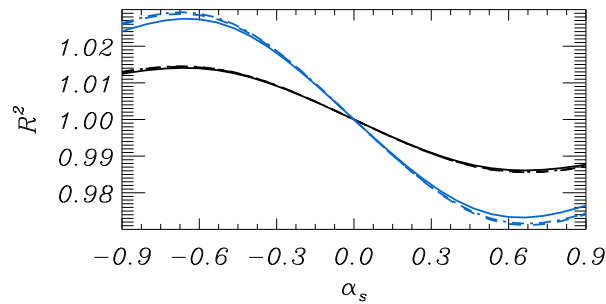}
\end{center}
\caption{Dependence of $R^2$ on convection ($\alpha_s$). Numerical data
(continuous line) are compared with Eq.~(\ref{eqR2}) (dashed-dotted line)
and Eq.~(\ref{eqR2CD}) (dashed line). $E_0=1.05E_0^c$ (dark lines), $
E_0=1.1E_0^c$ (light lines), $\gamma_p=\gamma_s=1$, $\Delta_p=-0.2$, $
\Delta_s=-0.5$, $a_p=0.5$, $a_s=1$.}
\label{fig05}
\end{figure}
The ratio between the intensities of the two critical modes also provides the
quantitative characterization of convection in the nonlinear symmetry breaking.
 The numerical and analytical
estimation of $R^2$  versus convection parameter $\alpha_s$,
displayed in Fig.~\ref{fig05}, are  again in very good agreement.
Finally, note on Fig.~\ref{fig05} the existence of extrema leading
to the most asymmetric configuration.

Let us now comment on the physics underlying the nonlinear symmetry
breaking induced by convection. The most relevant physical parameter
in the nonlinear interaction, above threshold, stems from the
difference in frequencies of oscillations of each mode ($\pm k_c$).
This difference results from the presence of convection and
disappears with it. Although, both traveling waves are propagating
in the direction of the convection, their phase rotation are no more
opposite. Hence, the two traveling modes interact with a time delay
with the pump. This gives rise to different gain from the pump
leading to the nonlinear symmetry breaking observed in the signal.
The energy transfer depends on two time scales and thus involves the
pump decay rate ($\gamma_p$) as can be seen from the expression of
$D$. We emphasize that in contrast with all previous studies dealing
with the weakly nonlinear dynamics of OPO near threshold, $\gamma_p
$ appears, for the first time, in the cubic Ginzburg-Landau model
because of the induced pump excitation phenomenon. This fixes the
parameter range of the pump decay rates leading to a nonlinear
symmetry breaking in the generated traveling waves. The stronger the
pump decay rate, the weaker the asymmetry is.  In
the limit of adiabatic elimination of the pump, no asymmetry exists
in the signal, consistently with the possibility to remove the
walk-off by a change of reference frame. We have performed
numerical simulations (not shown) with the same parameters as in
Fig.~\ref{fig02} except that $\gamma_p$ is decreased ten times. In
this case we have observed a vanishing asymmetry ($R^2 \simeq 1$)
with respect to the result of Fig.~\ref{fig02}.

\begin{figure}[tbp]
\begin{center}
\includegraphics[width=8cm]{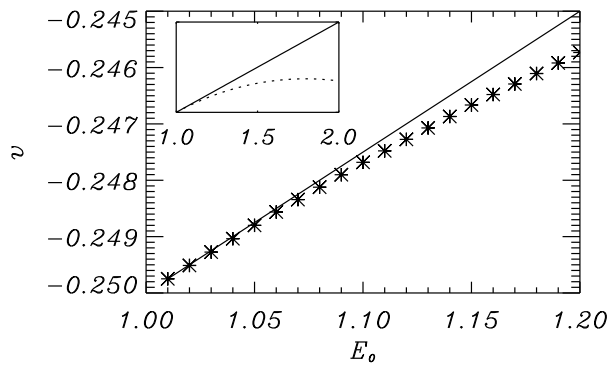}
\end{center}
\caption{Analytical velocity given by Eq.~(\ref{eqv}) (continuous line)
compared with numerical results (symbols) from the original model (\ref{eqdopo}).
In the inserted plot we compare the velocities far
from threshold. Same parameters of Fig.~\ref{fig02}.}
\label{fig09}
\end{figure}

The second important feature that results from the convection
induced nonlinear symmetry breaking concerns the propagation
velocity of the generated traveling waves. Indeed, the convection
effect on the signal is not only a translation of its transverse
profile at the convection velocity. An increase in the pump enhances
the action of convection in the nonlinear coupling between the
fields leading to the velocity variation with the pump intensity. So
that, if we set $\gamma _{s}\Omega =\omega _{cor}$, the corrected
frequency of the traveling waves is $\omega _{R}=\omega _{c}+\omega
_{cor}$. Therefore their actual velocity is given by
\begin{eqnarray}
v=\frac{\omega _{c}+\gamma _{s}\Omega }{k_{c}}=-\alpha _{s}\gamma
_{s}-\gamma _{s}\frac{\mathop{\rm Im}(\beta )}{k_{c}\mathop{\rm Re} (\beta )}
(\mu -1)
\label{eqv}
\end{eqnarray}
The above velocity expression shows that, in addition to the usual
convection velocity (the first term of the
right hand side) there is an excess velocity depending on the
convection but,
and most interestingly, it depends linearly on the incident pump above
threshold $(\mu -1)$. Figure  \ref{fig09} shows the
predicted deviation of the actual velocity from the velocity convection
by integrating the full nonlinear equations governing the DOPO
dynamics [Eqs. ~(\ref{eqdopo})] when increasing the pump $E_{0}$ till
twice the threshold. As can be seen from this figure, there is a  very
good quantitative agreement for a pump till $10\%
$ above
threshold. Only at threshold the nonlinear waves velocities coincide with
the convection velocity.

To summarize we have shown, in case of a degenerate optical parametric
oscillator, that convection
(walk-off) induced a nonlinear symmetry breaking in the traveling
waves. We have also demonstrated that near threshold this mechanism
is still described by a Ginzburg-Landau model with an original
dependence of the nonlinear self-coupling term upon convection. As a
result, nonlinear traveling waves are no more symmetrical and the
explicit analytical expressions of their intensities variations with
both convection and the distance from threshold are derived.
Moreover, convection leads to nonlinear phase modulations that give
rise to an interesting variation of the traveling waves velocity
with the distance from threshold. Besides the context of optics, our
results are relevant to many spatially nonlinear extended systems
with convection. For instance, in the context of hydrodynamics, the
competition between right- and left- propagating nonlinear waves in
the convective flow, generated by a horizontal thermal gradient,
leads to an experimental observation of the nonlinear symmetry
breaking. The broken symmetry has been evidenced
via the estimation of the variation of the
amplitude ratio of the right and left waves with the distance from
threshold \cite{Chiffaudel}.

This research was supported in part by The Programme International de
Coop\'eration Scientifique (PICS): France-Catalogne.
The Laboratoire de Physique des Lasers, Atomes et Mol\'ecules is Unit\'e de
Recherche Mixte du CNRS. The Centre d'\'Etudes et de Recherches Lasers et
Applications is supported by the Minist\`ere charg\'e de la Recherche, the
R\'egion Nord-Pas de Calais and the Fonds Europ\'een de D\'eveloppement
\'Economique des R\'egions.
R. Z. acknowledges financial support from the UK Engineering and
Physical Sciences Research Council (GR/S03898/01).

\end{document}